# The effects of citation-based research evaluation schemes on self-citation behavior

**Authors:** Giovanni Abramo[1]*, Ciriaco Andrea D'Angelo[2,1], Leonardo Grilli[3]

**Affiliations:**

[1] Laboratory for Studies in Research Evaluation, Institute for System Analysis and Computer Science (IASI-CNR). National Research Council, Rome, Italy
ORCID: 0000-0003-0731-3635 - giovanni.abramo@uniroma2.it

[2] University of Rome "Tor Vergata", Dept of Engineering and Management, Rome, Italy
ORCID: 0000-0002-6977-6611 - dangelo@dii.uniroma2.it

[3] University of Florence, Dept of Statistics, Computer Science, Applications "G. Parenti", Florence, Italy
ORCID 0000-0002-3886-7705 - grilli@disia.unifi.it

*corresponding author*

**Abstract**
We investigate the changes in the self-citation behavior of Italian professors following the introduction of a citation-based incentive scheme, for national accreditation to academic appointments. Previous contributions on self-citation behavior have either focused on small samples or relied on simple models, not controlling for all confounding factors. The present work adopts a complex statistics model implemented on bibliometric individual data for over 15,000 Italian professors. Controlling for a number of covariates (number of citable papers published by the author; presence of international authors; number of co-authors; degree of the professor's specialization), the average increase in self-citation rates following introduction of the ASN is of 9.5%. The increase is common to all disciplines and academic ranks, albeit with diverse magnitude. Moreover, the increase is sensitive to the relative incentive, depending on the status of the scholar with respect to the scientific accreditation. A further analysis shows that there is much heterogeneity in the individual patterns of self-citing behavior, albeit with very few outliers.

# 1. Introduction

With the arrival of the knowledge-based economy there has been a global rush to improve the effectiveness and efficiency of national research systems. To that aim, an increasing number of governments have introduced or strengthened incentive schemes designed to increase research productivity in universities and public research organizations. With this, comparative research assessment has also flourished, supported by evaluative scientometrics.

The increasing resort to evaluative scientometrics has led scholars and policy makers to inquire into possible collateral effects on the strategic management of the research institutions under assessment, and on the behavior of individual researchers, particularly when institutions deploy incentives at the individual level, through monetary rewarding systems, access to resources, recruitment, career advancement, etc. (Geuna & Martin, 2003; Tonta, 2017; Moher et al., 2018).

In 2017, the Journal of Informetrics dedicated a special section to the effects of metrics-based funding systems on scientists' behavior (Volume 11, Number 3). The debate opened with a discussion paper by van den Besselaar, Heyman, and Sandström (2017), objecting to Butler's pioneering works on the behavioral effects of the Australia funding formula (Butler, 2003a; 2003b), and in particular criticizing her scarce attention to the impact of the scientist's production. Several other scholars contributed to the debate (Aagaard & Schneider, 2017; Gläser, 2017; Hicks, 2017; Martin, 2017).

According to "rational cheaters" model, employees anticipate the consequences of their actions and adopt opportunistic behavior when the marginal benefits exceed marginal costs (Nagin, Rebitzer, Sanders, & Lowell, 2002). Proceeding from such theory, several scholars have hypothesized and investigated perverse effects on the behavior of researchers and institutions, caused by the increasing resort to metrics in evaluation systems (Fang, Steen, & Casadevall, 2012; Haustein S., & Larivière V., 2015; De Rijcke, Wouters, Rushforth, Franssen, & Hammarfelt, 2017). Among these, evidence is found for a range of scientific misconducts, such as multiplication of irrelevant publications, plagiarism, self-plagiarism and scientific fraud (Hazelkorn, 2010; Honig & Bedi, 2012; Edwards & Roy, 2017), discouragement of research diversification, and of interdisciplinary or innovative research (Hicks, 2012; Rafols, Leydesdorff, O'Hare, Nightingale, & Stirling, 2012; Wilsdon, 2016; Abramo, D'Angelo & Di Costa, 2018).

In recent years, the "publish or perish" syndrome (Fanelli, 2010; Van Dalen & Henkens, 2012; Neill, 2018) has evolved into the more sophisticated "top cited or perish" syndrome (Chandler, Barry, & Clark, 2002; Gill, 2009; Sà, Kretz, & Sigurdson, 2013). Researchers and managers are immediately aware when countries introduce national research assessment exercises using citation metrics, and attentive to the fact that funding agencies and tenure committees use citations in decision making. In some nations, authors are even paid bonuses for highly cited articles (Abritis & McCook, 2017; Stephan, Veugelers, & Wang, 2017; Tonta & Akbulut, 2019). This has led observers and scholars to criticize and question the potential phenomenon of self-citation gaming by scientists (Biagioli, 2016; Scarpa, Bianco, & Tagliafico, 2018; Seeber, Cattaneo, Meoli, & Malighetti, 2019; Baccini, De Nicolao, & Petrovich, 2019; D'Antuono & Ciavarella, 2019).

The present work contributes to this stream of research, with the objective of verifying if and to what extent scientists increase self-citation rates when strongly tempted by incentivizing schemes. In case they do, we will inquire into the diffusion or concentration of the phenomenon, and extend the assessment to the propensity for increasing self-citations by discipline, gender, academic rank and territory.

One of the strengths of the study is its approach to the measurement of self-citation. Almost all of the previous studies on self-citation behavior have adopted a simplified approach, taking the publication as the base unit of analysis. According to this approach, self-citation is represented by the sharing of at least one co-author between two publications, cited and citing. Instead, our approach is to consider the author as the basic unit of analysis, and therefore to identify self-citation as the case of an author of a citing paper who is also the author of the cited paper, but not the case of co-authors who cite it.

The realization of the research project requires resolution of several obstacles in bibliometric operationalization, as well as assumptions that become ever more inevitable with larger scale of analysis. In formulating the methodological approach we pay particular attention to motivating our choices and specifying the limits in the operationalization of the measures, providing the reader with all the elements necessary for evaluation of the robustness and reliability of the results, for their comparison with previous studies, and for facilitation of replication in other contexts. In this way, it should be easier for our readers overcome the difficulties in comparing to other studies, given the common resort to different methodologies and indicators.

From the outset, we can expect that the opportunistic response to incentive schemes would be more or less pronounced, and concentrated in a few or widespread among many, depending on: i) the potential benefits, ii) the costs of implementing the gaming, iii) the risk of being discovered, iv) the penalties in case of discovery, and finally, v) the ethical sense of the scientific community under observation. In any assessment of self-citation gaming or interpretation of the results we must therefore consider the reference context, and the generalization or international comparison of results would require the utmost caution.

We should immediately note that an increase of self-citation rates does not necessarily imply gaming, i.e. illegitimate self-citing. Among other reasons, such increases could be the result of the authors scrupulously crediting their true attention to their own works, knowing that will be subject to citation-based evaluations.

To address our objective, we contrast the self-citation behavior of professors in Italy before and after a significant reform to the university recruitment system, intended, among others, to halt perceived cases of poor and mediocre candidates from accessing the ranks of associate and full professor. The reforms imposed that for these ranks, all prospective candidates would have to complete of a preliminary phase of National Scientific Accreditation (ASN). In the hard sciences in general, this required achievement of threshold values in up to three bibliometric indicators of research performance, two of which were citation-based.

In the next section, we delve into self-citation behavior and the difficulties of measuring self-citations, with the intent of delineating the "area of self-citation" that can be considered to result solely from opportunistic behavior. We advance this task further in Section 3, through a review of literature on self-citation behavior. In particular, we analyze the methodological approaches of previous studies, and the reliability of their findings. In Section 4, we develop our framework of investigation, presenting the field of observation and methods. Section 5 presents the results, and Section 6 the authors' considerations and recommendations.

## 2. Self-citation: behavior and measurement

The approach of evaluative citation-based scientometrics is conceptualized on the basis of phenomena theorized and studied by the sociologists of science. In a narrative review of studies on the citing behavior of scientists, Bornmann and Daniel (2008) analyze the motivations that lead scientists to cite the work of others. The findings show that "citing behavior is not motivated solely by the wish to acknowledge intellectual and cognitive influences of colleague scientists, since the individual studies reveal also other, in part non-scientific, factors that play a part in the decision to cite." Nevertheless, "there is evidence that the different motivations of citers are not so different or randomly given to such an extent that the phenomenon of citation would lose its role as a reliable measure of impact."

In the sociology of science, two different theories of citing behavior have been developed: the normative theory and the social constructivist view. The first, based on a milestone work of Robert Merton (1973), affirms that scientists, through the citation of a scientific work, recognize a credit towards a colleague whose results they have used, meaning that the citations represent an intellectual or cognitive influence on their scientific work.

A second view on citing behavior is based on constructivist theory advanced in the sociology of science (Latour & Woolgar, 1979; Knorr-Cetina, 1981). The social constructivist approach contests the base assumptions of normative theory, and so challenges the validity of evaluative citation analysis. Constructivists argue that "scientific knowledge is socially constructed through the manipulation of political and financial resources and the use of rhetorical devices" (Knorr-Cetina, 1991), meaning that citations would not necessarily be linked in consequential manner to the scientific contents of the cited article.

Tahamtan, Safipour Afshar, and Ahamdzadeh (2016), and Tahamtan and Bornmann (2018) provide reviews of the theoretical literature and technical studies on citing behavior.

Self-citations represent a particular subclass of citations. The recourse to self-citing is legitimate when there are good scholarly reasons for self-citing, first of all, recognizing a credit towards one's own work, as one would do towards others. Authors would also wish to cite previous, relevant work to avoid allegations of self-plagiarism or redundant publication (Garfield, 1979; Pichappan & Sarasvady, 2002). Self-citing is instead illegitimate when the cited publications do not represent a true contribution to the scholarly content of the new publication. The inclusion of such self-citations could be suspect as a mechanism of increasing citations, whose counting would then misrepresent the importance of the specific work and also the publishing journal, and ultimately distort any evaluations.

There would seem to be three main reasons for authors to use self-citation in an attempt to increase their citation rate, in line with social-constructivist theory on manipulation and rhetorical device.

One reason for inflating self-citations would be the self-serving tendency to personal gratification, as well as self-enhancement and self-promotion, in the struggle for visibility and scientific authority (Hyland, 2003; Brysbaert & Smyth, 2011).

There would also be the strategic aim of using self-citing as an advertising and promotional tool, especially for recent publications, to increase their citation by others (Van Raan, 2008). It has been estimated that "each additional self-citation increases the number of citations from others by about one after one year, and by about three after five years. Moreover, there is no significant penalty for the most frequent self-citers – the effect of self-citation remains positive even for very high rates of self-citation" (Fowler & Aksnes, 2007).

A third reason would be to exploit the opportunity of incentive mechanisms that reward high citation rates, in particular when evaluative scientometrics could be applied to inform decisions concerning the self-citer.

Given that our objective is to assess the effects on self-citation rates of an accreditation policy relying on evaluative scientometrics, we must then distinguish such opportunistic self-citation behavior, first of all from legitimate uses, but also from the other illegitimate forms of self-citation. Indeed, any investigator, including when proposing perverse effects, must take steps to account for these confounding factors, to avoid bias and misinterpretation of the results on analysis of self-citation.

Whatever the motive, the mechanisms of illegitimate self-citing can be direct or indirect. Self-citations are "direct" when the (co)authors of a paper cite previous papers (co)authored by themselves. Self-citations are "indirect" when: i) "coercive induced", where a manipulator, such as a scientific advisor, article or grant-application reviewer, induces other scientists to cite their own works, even though these do not contribute to the citing articles (Thombs, 2015; Ioannidis, 2015); or ii) scholars collude to cross-cite one another's works (Sala & Brooks, 2008). Both coercive induced and "collusive" self-citation, also called "cross-citation", are very difficult to discover and would not be caught in large-scale bibliometric analyses. This problem also presents limits on most studies concerning self-citation behavior itself.

In investigating direct self-citation behavior, a critical issue is the definition of self-citation, and the consequent operationalization of measurement. The literature provides two definitions. The first of these applies common sense, beginning with the conception of the author: a citation is defined as a self-citation whenever the author of a cited paper (in co-authorship or not) is also a (co)author of the citing paper. Operationally, this definition entails the author as the base unit of analysis. For convenience, we will call this approach "author-based".

A second approach instead begins from the publication: a citation is defined as self-citation whenever the set of co-authors of the citing paper and the one being cited are not disjoint, that is, if the two sets share at least one person (Snyder & Bonzi, 1998). This definition, however, extends beyond the real meaning of self-citation, since an individual author might result as a serial self-citer even if they never self-cited their own works. This would occur when, among the clusters of publication co-authors, there are serial self-citers: in that case, one author would "import" self-citations from the actions of others. We call this approach "publication-based".

Given the extension of the meaning in taking the publication as base unit, we would expect that this will result in the observation of higher self-citations as a proportion of total citations. Indeed, in a macro-study of more than half a million citations of articles by Norwegian scientists listed in the Science Citation Index, Fowler and Aksnes (2007) found that the publication-based approach attributes remarkably higher self-citation rates (21%) than the one taking the author as base analytical unit (11%).

Scholars frequently state that the choice of publication-based approach is intended to unveil collusive indirect cross-citation. In most cases, though, we suspect that the reason is that it affords much greater operational ease in the detection and measurement of self-citations. The adoption of the author-based definition, in fact, demands the initial disambiguation of all author names, requiring lengthy manual work or the development of disambiguation algorithms. The publication-based definition instead offers much simpler operationalization, since it requires only the identification of the intersection of the author sets of cited and citing publications, as observed in the bylines. This would explain why very few studies have taken the author as base unit of analysis (Fowler & Aksnes, 2007; King, Bergstrom, Correll, Jacquet, & West, 2017; Seeber, Cattaneo, Meoli, & Malighetti, 2019).

Our personal belief is that the author-based approach is the appropriate one, unless we imagine that the global scientific community somehow encompasses vastly interconnecting cartels of co-authors, busily engaged in cultivating cross-citations, and that this is the reality we wish to investigate. In fact, taken apart, the aim of investigating collusive cross-citing would be better performed using techniques of co-authorship network analysis (González-Teruel, González-Alcaide, Barrios, & Abad-García, 2015; Zhao & Strotmann, 2015).

What is certainly clear is that when preparing an analysis, interpreting and comparing published results, or generally examining the literature on self-citations, we must first distinguish the choices of base definitions.

Having chosen a conceptual definition, and proceeding to method, self-citations can then be "counted" in two ways, synchronously (looking back), or diachronously (looking forward) (Lawani, 1982; Glänzel, Thijs, & Schlemmer, 2004; Costas, Van Leeuwen, & Bordons, 2010). Taking a given paper, the synchronous approach counts the author's self-citations to the publications listed in the references. In theory, the numerosity observed is deterministic. This approach, however, entails authorship disambiguation of all previous publications back to the date of the author's very first, which could have been decades earlier, representing a formidable task. In practice then, the citable publications search is often cut off at a chosen historic date, making the self-citation count a prediction.

Conversely, taking a given paper, the diachronous approach "looks forward", counting the self-citations received, however in this approach the observation of numerosity depends on the choice of citation time window. This method therefore also offers a prediction, the more accurate the longer is the citation window.

## 3. The literature on self-citation behavior

Given the central role of citations in evaluative scientometrics, and the potentials of distortion from illegitimate self-citations, practitioners have been under increasing pressure to formulate indices for the identification and possibly the sanction of extreme self-citers. Most of the indicators advanced have been "self-citation rates", with different choices of numerator and denominator depending on the underlying conception. The numerator of "self-citations" is common to all, but even this requires the definition of self-citation. Then, with the choice of different denominators, we have self-citation rates such as: the ratio of author's total self-citations to total number of received citations; ii) ratio of total self-citations to the individual's total publications; iii) the average of the ratios of self-citations to citations in each of the author's papers.

There is also the very simple proposal of only the average number of self-citations per paper, and the highly creative "$s$-index", structured similarly to the famed $h$-index. In this version, $s$ is the number of papers with at least $s$ self-citations by the author (Flatt, Blasimme, & Vayena, 2017).

Recently, practitioners have published the first ranking lists of researchers by self-citation rates. Ioannidis, Baas, Klavans, and Boyack (2019) measured the ratio of self-citations to total received citations for the 100,000 global top-cited researchers. Unfortunately, the definition of self-citations is "publication-based", meaning that the presence of any co-author shared between the citing and cited publications counts as a self-citation, with consequences leading to evident distortions of ranking. For example, perusing the data provided by the authors, we see the case of Sonia Del Prete, researcher at the Institute of Biosciences and Bioresources of the Italian National Research Council. In 2017, her articles published in that same year received 456 total citations, of which 92% would be self-citations. Of the 100,000 top-cited researchers, only eight show a higher self-citation rate. In reality, a manual "author-based" verification of the data reveals that Sonia Del Prete's true self-citations counted less than 25%, and that 97% of her publications were co-authorships involving a much more prolific colleague (from the University of Florence), whose true self-citations are over 56%. This is an evident case of "importing" the self-citations of others, due the publication-based approach. In such applications, one of the very unfortunate consequences is the risk of exposing "innocent" scientists to pillory within and beyond the scientific community.

It must be said that the authors of the new ranking list cautioned against inventing threshold self-citation rates, distinguishing "honest" from "unethical" researchers, or against any inference that could lead to vilification of individuals for their personal self-citation rates.

The new study, however, does stimulate thought of further lines of in-depth investigation, possibly leading to identification of extreme direct self-citers or cross-citation cartels, as well as the contexts of their occurrence.

Our position, however, is that using self-citation rate indicators to compare citation behavior at the individual level is open to misleading conclusions and grave risks. There could be many legitimate explanatory factors for behaviors in which individual scientists self-cite more than others. The literature provides useful suggestions on such potential factors, and each of these requires consideration in the adoption of inferential models.

There have been several studies on self-citations examining a limited number of disciplines and taking a small number of papers, but very few large-scale studies and even less considering the underlying behaviors. In the following, we review what is available, searching for information useful in formulating the statistical model needed to answer our specific research question: do scientists increase self-citation rates when strongly tempted by incentivizing schemes, and if so, to what extent?

Researchers tend to self-cite recent publications more than older ones, and poorly cited papers more than highly cited (Aksnes, 2003). Aksnes also finds a positive correlation between the number of self-citations and the number of authors of the publications (Aksnes, 2003). It is also known that internationally co-authored publications are more highly self-cited than domestic (Adams, 2013; Kumar, Rohani, & Ratnavelu, 2014), and this holds true for the publications with Italian co-authors (Abramo, D'Angelo, & Murgia, 2017) whether or not the Italian self-citations are included (Abramo, D'Angelo, & Di Costa, 2020).

Researchers differ in the diversification of their research activity (Abramo, D'Angelo, & Di Costa, 2017; Stirling, 2007), and we would expect that, all else equal, those pursuing a specialized line will also be more likely to self-cite.

Different disciplines show significant differences in self-citation rates: high in the sciences, lower in the social sciences, and lowest in the humanities (Snyder & Bonzi, 1998; Aksnes, 2003). Differences can also occur between the research fields within disciplines. This suggests the need to differentiate at the field level, for assessment of propensities to engage in opportunistic self-citation.

It seems that men tend to cite their papers 70 percent more than women (King, Bergstrom, Correll, Jacquet, & West, 2017; Mishra, S., Fegley, B.D., Diesner, J., & Torvik, V.I., 2018). They also produce more papers, and so have more to cite (Sugimoto, Lariviere, Ni, Gingras, & Cronin, 2013; Elsevier, 2020), and this certainly holds true in Italy (Abramo, Aksnes, & D'Angelo, 2020; Abramo, D'Angelo, & Caprasecca, 2009). For Denmark, Nielsen (2016) did not detect statistically significant differences in self-citation rates between men and women, except in the medical sciences. This suggests that in examining behavioral response to the ASN, we should also consider gender differences.

It seems that citations per paper vary across the academic ranks of authors (Ventura & Mombrù, 2006). In Italy, the average of citations per paper by full professors is slightly higher than that for associate professors, and remarkably higher than for assistant professors (Abramo, D'Angelo, & Di Costa, 2011). Given this, we should also inquire to what extent self-citation behavior varies across academic ranks.

Organizations are characterized by distinct identities, and core values, beliefs and ethical codes that serve in guiding employee behavior (Collins & Porras, 1991). We could therefore expect that across universities there would be different tendencies to artificially increase self-citations. There are can also be distinct cultural, social, and economic disparities between territories, as occur between northern and southern Italy (Cafagna, 1989; Daniele, 2015), and in the Italian case these differences are also accompanied by a remarkable gap in research performance (Abramo, D'Angelo, & Rosati, 2016). Such territorial differences could represent further influences on behavior.

The literature on the perverse effects of evaluative scientometrics on scientists' behavior in general is quite rich, however there is very little concerning the effects of specific incentive schemes on self-citation. We have found only four studies strictly focused on the topic and, quite strikingly, all concern the Italian field of observation.

Scarpa, Bianco, and Tagliafico (2018) inquired into the effects on publication and self-citation rates of the national Research Quality Evaluation (VQR) exercise and the ASN, introduced respectively in 2011 and 2012. The study was limited to professors in engineering (7151 in total). For each professor, over the 2006-2016 period, the authors measured the yearly self-citation rate, defined as total self-citations divided by total citations received. Based on this indicator, the descriptive analysis showed a conspicuous increase of the average self-citation rate after the introduction of the citation-based evaluations. The authors concluded that "the harmful influence of the introduction of evaluation exercises is clearly visible by analyzing the authors' self-citation habits. This datum pushes for the use of impact indices which are scarcely sensible to such a phenomenon." The study is subject to important limits, including: i) the lack of inferential analysis considering factors other than the desire to inflate one's own citations, which might justify the increase in self-citations; ii) the consideration of only one discipline.

Seeber, Cattaneo, Meoli, and Malighetti (2019) conducted a similar investigation on a total of 866 professors belonging to the four research fields of Economic and managerial engineering, Genetics, Psychiatry, and Applied economics (the latter as a control field). The analysis applied descriptive and inferential statistics to portray the time evolution of self-citations between 2002 and 2014. The authors concluded that "introduction of a regulation (ASN in 2012) that links the possibility of career advancement to the number of citations received is related to a strong and significant increase in self-citations among scientists who can benefit the most from increasing citations, namely assistant professors, associate professors and relatively less-cited scientists". In our own view, while the investigation applied a rigorous empirical approach, the scope (three scientific fields out of an identified 109, for Italy) was too narrow to arrive at such general conclusions.

Baccini, De Nicolao, and Petrovich (2019) also studied the impact of the ASN on self-citation behavior. They conceived a country-level indicator, "especially designed to be sensible to the effects of both the opportunistic use of author self-citation and the creation of citation clubs". The indicator, named "inwardness", is defined as the ratio of the total number of self-citations by a country to the total of all citations received by the same country. The authors then compared the 2000-2016 evolution of inwardness in Italy with that of the other G10 states. Observing a significant increase of inwardness after the launch of the ASN, relative to the evolution of other countries, they concluded that the Italian scientific community had adopted opportunistic strategies as a response.

This "inwardness" study has attracted wide attention (Guglielmi, 2019; Huet, 2019; Magnani, 2019; Van Noorden, 2019; Singh Chawla, 2019), however it suffers limits in methodology and from the indicator itself. Among others: i) national self-citations are defined as citations from within the country, but also from the countries of any co-authors, regardless of whether the foreign citing authors were truly the co-authors of the cited paper.[1] One could conclude a global scientific community capable of organizing behavior to the level of clusters of countries. ii) Even when self-citations decline, the country's inwardness score could rise, in the simple case that the denominator, total received citations, decreases more. iii.) The ASN uses citation-based indicators for evaluation of academic candidates only in the sciences, and for these could be an incentive to manipulate citations. However, the study also applies the inwardness indicator to the social sciences, arts and humanities, and to the publications of a broad community very little concerned with academic accreditation, including private-sector researchers and those in the public but non-academic sphere. iv) The study lacks inferential analysis of confounding variables.

Subsequent to the above-noted study by Ioannidis et al. (2019), proposing the ranking list of the 100,000 global top-cited researchers, D'Antuono and Ciavarella (2019) used the same dataset to compare the self-citation rates of Italian top scientists to those of other national scientists. Based on the results, the authors concluded that "the problem of self-citations, in relation to this scientific elite, is not significant in Italy."

**4. National Scientific Accreditation: an incentivization scheme[2]**

Scholars have dedicated considerable attention to problems of fairness in academic appointments, largely concerning gender and minority discrimination (Trotman, Bennett, Scheffler, & Tulloch, 2002; Price et al., 2005; Cora-Bramble, 2006; Stanley, Capers, & Berlin, 2007; Zinovyeva and Bagues, 2015; De Paola & Scoppa, 2015). Discriminatory phenomena are more likely when evaluations use non-transparent criteria (Rees, 2004; Ziegler, 2001; Husu, 2000), which can enable powerful professors to engineer the selection of candidates through cooptation mechanisms (Van den Brink, Benschop, &

---
[1] To exemplify, all citations from USA publications to an Italy-USA co-authored publication are considered as Italian self-citations.
[2] The description that follows includes text from a previous publication (Abramo, & D'Angelo, 2015).

Jansen, 2010; Husu, 2000). Cooptation is often coactive with favoritism, which has been intensively examined abroad (Aydogan, 2012; Martin, 2009; Zinovyeva and Bagues, 2015; Li & Tang, 2019) and in Italy, where the phenomenon seems endemic (Abramo, D'Angelo, & Rosati, 2014, 2015, 2016; Perotti, 2008; Zagaria, 2007).

For decades, both the popular and academic media have published criticism on the nationally operated competitions for Italian academic positions. As early as 2001, Gerosa remarked that "the Italian word *concorso* has gained international note for its implications of rigged competition, favoritism, nepotism and other unfair selection practices." Governments made countless attempts to fight favoritism, by changing the rules and procedures of the academic competitions, but with modest success.

Law 240 of 2010 set out policy and criteria for broad reforms of the university system. In the area of recruitment, it established that competitions for positions would be limited to candidates previously qualified under the new ASN scheme. In 2011 and 2012, a "Regulation on awarding of national scientific accreditation"[3] and "Regulation of criteria and parameters for evaluation of candidates and verification of committee member qualifications"[4] elaborated the structures and operation of the ASN.[5] The dates of these administrative acts are important in establishing the framework of our empirical investigation.

The law and regulations established a framework of Accreditation Committees for operation of the competitions, and importantly, directed that these would be informed by bibliometric indicators of research performance, to be developed by the National Agency for Evaluation of the University System (ANVUR). For positions of associate and full professor, passing the threshold values of one, two or three (depending on the competition sector) bibliometric indicators[6] became a prerequisite for accreditation, and thus for career advancement – clearly establishing a "strongly tempting" incentivizing scheme, relevant to our current research question.

In detail, according to the regulatory framework, every two years the MIUR must issue a call for appointments to 184 Accreditation Committees, with one committee for each Competition Sector (CS). The CSs derive from the aggregation of the so-called Scientific Disciplinary Sectors (SDSs, 370 in all), which serve in general for the classification and administration of all Italian university faculties: under this system, each professor is unequivocally classified as practicing in a specific SDS. The SDSs, and so also their aggregations in CSs, are grouped in 14 University Disciplinary Areas (UDAs). Of the 184 CSs, ANVUR (the Italian evaluation agency) classifies 109 as "bibliometric", and the remaining 75 as "non-bibliometric". For the bibliometric CSs, applicant members of the accreditation committees are chosen randomly from all full professors categorized in the CS who pass all three bibliometric thresholds.

The call for candidates to accreditation was instead originally issued on an annual basis, and more recently quarterly. Applicants must apply for specific CSs and academic ranks, however there is no limit on the number of CSs requested, and candidates can request accreditation to both associate and full professor at the same time. The candidatures were to be accompanied by a curriculum vitae and a list of the individual's

---

[3] Presidential decree 222 of 14/09/2011.
[4] Ministry of Education, Universities and Research (MIUR) Ministerial decree 76 of 07/06/2012.
[5] The description that follows includes text from a previous publication (Abramo, & D'Angelo, 2015).
[6] Initially the regulations required achievement of thresholds in all the three indicators, but after heated debate this was modified in consideration of the conditions of different sectors.

degrees and scientific publications.

For the bibliometric CSs, the list of publications served for the calculation of the three indicators.[7] For the candidates to the 2012 accreditations, they were:[8]

- number of articles in journals over the period 2002-2012, with normalization in the case of academic seniority less than 10 years (calculated from the first year of publication);
- number of citations received for the overall scientific production, normalized for the academic seniority of the candidate;
- contemporary h-index (Sidiropoulos, Katsaros, & Manolopouloset, 2007) of the overall scientific production.

The above indicators were calculated on the basis of the publications indexed in two major bibliographic repositories: Scopus and Web of Science (WoS). In the case of publications indexed in both sources, the value of citations assigned was the higher of the two.

Beginning from a database of all the publications entered by Italian professors over the course of 2012, ANVUR proceeded to calculate the different indicators for full and associate professors and then published the median values, to serve in the selection of both the committee members and their subsequent evaluations of the candidates.

Table 1 provides the descriptive statistics for the 2012 call for accreditations. The response was over 59,148 applications (55.7% already on faculty staff), more than one sixth of which were for Medicine (9987). The requests for full professor accreditation were 30.5% of the total.

*Table 1: Statistics for the 2012 National Scientific Accreditation (from base data published on the ASN pages of the MIUR website http://abilitazione.miur.it/public/candidati.php?sersel=50&)\**

| UDA | CS | Of which "bibliometric" | Applications | Of which for full professor |
|---|---|---|---|---|
| 01 - Mathematics and computer science | 7 | 7 | 2,492 | 911 (36.6%) |
| 02 - Physics | 6 | 6 | 4,372 | 1,451 (33.2%) |
| 03 - Chemistry | 8 | 8 | 2,344 | 695 (29.7%) |
| 04 - Earth sciences | 4 | 4 | 1,231 | 400 (32.5%) |
| 05 - Biology | 13 | 13 | 6,244 | 1,690 (27.1%) |
| 06 - Medicine | 26 | 26 | 9,987 | 3,298 (33.0%) |
| 07 - Agricultural and veterinary sciences | 14 | 14 | 2,093 | 650 (31.1%) |
| 08 - Civil engineering and architecture | 12 | 7 | 3,599 | 1,027 (28.5%) |
| 09 - Industrial and information engineering | 20 | 20 | 4,535 | 1,573 (34.7%) |
| 10 - Ancient history, philology, literature, art | 19 | 0 | 6,324 | 1,718 (27.2%) |
| 11 - History, philosophy, pedagogy, psychology | 17 | 4 | 5,909 | 1,491 (25.2%) |
| 12 - Law | 16 | 0 | 3,037 | 887 (29.2%) |
| 13 - Economics and statistics | 15 | 0 | 4,853 | 1,755 (36.2%) |
| 14 - Political and social sciences | 7 | 0 | 2,128 | 515 (24.2%) |
| Total | 184 | 109 | 59,148 | 18,061 (30.5%) |

\* *Subject to regulation, the ministry withdrew most data 60 days after the committee decisions, leaving only the publication of the lists of successful candidates.*

---

[7] For the "non-bibliometric" CSs, the indicators were not based on citations.
[8] For applicants to positions as committee members, the indicators were number and impact of overall scientific production, without normalization for years of academic seniority.

## 5. Research hypotheses, data and methods

For accreditation of assistant and associate professors hoping to advance in bibliometric CSs, and for full professors seeking selection to the relative Assessment Committees, two of three ASN bibliometric indicators are based on citations. For these academics, the ASN scheme clearly provides a potential incentive to inflate one's own citations. The incentive could extend to non-academics hoping to become professors, however we limit our analysis to the academics, since only for these do we have the necessary public database listing their names, disciplinary sectors, rank, etc.

From our research question, we subsume two hypotheses:

**Hypothesis 1**. Assistant and associate professors in bibliometric CSs, eager to progress in their careers, will increase their self-citation rates following introduction of ASN.

**Hypothesis 2**. Full professors, and associate professors accredited under the first ASN procedure, being subject to a weaker incentive, will increase their self-citation rates to a lesser extent.

To test the hypotheses, we analyze the self-citation behavior of professors in two five-year periods prior and subsequent to the introduction of ASN.[9] The accreditation criteria and indicators were first published in June 2012, so it is reasonable to expect the first appearances of any self-citation gaming in publication would be after 1 January 2013.[10] We then analyze the yearly number of self-citations as indexed in WoS, by Italian professors in bibliometric CSs, across the 5 years prior to and after that date, i.e. 2008-2017. Any abrupt increases in self-citations after 2012, all else equal, would signal possible opportunistic behaviors by Italian faculty.

### 5.1 Dataset

According to the database of Italian professors maintained by the MIUR,[11] at the end of 2012 there were 57,400 professors on faculty at Italian universities, of which 35,242 belonged to bibliometric CSs.

We use the author name disambiguation algorithm developed by D'Angelo, Giuffrida, and Abramo (2011) for construction of the bibliometric dataset, based on the coupling of the publications extracted from the Italian National Citation Report (I-NCR) and the MIUR database. This algorithm[12] assigns an I-NCR publication to a given professor if the latter:
- has a name compatible with one of the authors of the publication;
- belongs to one of the recognized universities in the list of addresses indicated by the authors of the publication;

---

[9] The ongoing ASN calls have followed each other in time at different intervals.
[10] It would be extremely difficult to write a paper with inflated self-citations, pass refereeing and achieve publication, in less than six months.
[11] http://cercauniversita.cineca.it/php5/docenti/cerca.php, last accessed on 23 September 2020.
[12] The harmonic average of precision and recall (F-measure) of authorships, as disambiguated by the algorithm, is around 97% (2% margin of error, 98% confidence interval).

- belongs to an SDS compatible with the subject category of the publication;
- held a tenured faculty position at 31 December of the year prior to that of the publication.

The I-NCR available to us is composed of all 2001-2017 publications[13] indexed in the WoS core collection, with "Italy" as country affiliation of at least one author. The dataset is composed of the Italian professors (15,037 in all), in role in a bibliometric CS in each year of the period 2001-2017. Table 2 shows the breakdown of this database by disciplinary area.

*Table 2: Dataset of analysis*

| UDA* | No. of CSs | No. of SDSs | No. of professors** | | | |
|---|---|---|---|---|---|---|
| | | | Full | Associate | Assistant | Total |
| 1 | 2 | 10 | 649 | 598 | 287 | 1,534 |
| 2 | 4 | 8 | 347 | 521 | 202 | 1,070 |
| 3 | 4 | 11 | 445 | 660 | 350 | 1,455 |
| 4 | 1 | 12 | 153 | 226 | 112 | 491 |
| 5 | 9 | 19 | 760 | 814 | 508 | 2,082 |
| 6 | 12 | 50 | 1,312 | 1,355 | 965 | 3,632 |
| 7 | 9 | 30 | 543 | 626 | 290 | 1,459 |
| 8 | 2 | 9 | 310 | 281 | 69 | 660 |
| 9 | 8 | 42 | 1,123 | 994 | 250 | 2,367 |
| 11 | 1 | 8 | 159 | 108 | 20 | 287 |
| Total | 52 | 199 | 5,801 | 6,183 | 3,053 | 15,037 |

\* 1 - Mathematics and computer science, 2 - Physics, 3 - Chemistry, 4 - Earth sciences, 5 - Biology, 6 - Medicine, 7 - Agricultural and veterinary sciences, 8 - Civil engineering, 9 - Industrial and information engineering, 11 – Psychology
\*\* Counts refer to rank and discipline as of 31/12/2012.

Bibliometric data of each professor are arranged as a 10-year panel. The structure is unbalanced since the self-citation rate is undefined in any year when an author has zero publications. It results that 6,224 authors have observations on all 10 years and the rest have at least one year with missing value. Overall, 120,615 professor-years are available.

The 15,037 professors produced a total of 563,457 publications, distributed per UDA and year as indicated in columns 2 and 5 of Table 3. Note that the structure of the dataset is unbalanced, since the self-citation rate is undefined in any year when a professor has zero publications.

The publications authored contain a total of 1,062,559 self-citations, with an average of 1.74 self-citations per publication (self-citation rate) in the period 2008-2012, and 2.00 in the period 2013-2017 (columns 4 and 7 of Table 3).

In the first period, the UDA with the highest average self-citation rate was Physics (3.18), followed by Chemistry (2.80) and Biology (2.24). In all other UDAs, the average rate was less than 2, with the minimum at 0.86, in Industrial and information engineering. In the second period, in all disciplines, the average self-citation rate is higher. The UDAs with highest and lowest rates remain the same (Physics at 3.87; Industrial and information engineering at 1.07). The largest increase was in Civil engineering (from 1.07 self-citations per publication in the first period to 1.61 in the second, i.e. +50%) followed by Agricultural and veterinary sciences (1.67 vs 1.25, i.e. +34%).

---

[13] Including journal articles, reviews, letters and conference proceedings; excluding editorials, meeting abstracts, reprints, and any other document types not embedding research results.

*Table 3: Publications and self-citations of professors in the dataset*

|  | 2008-2012 | | | 2013-2017 | | |
|---|---|---|---|---|---|---|
| UDA* | Publications | Self-citations | Self-citations/ publications | Publications | Self-citations | Self-citations/ publications |
| 1 | 13798 | 13902 | 1.01 | 15513 | 18511 | 1.19 |
| 2 | 32474 | 103192 | 3.18 | 38475 | 148897 | 3.87 |
| 3 | 29450 | 82383 | 2.80 | 33098 | 101377 | 3.06 |
| 4 | 4955 | 9231 | 1.86 | 6628 | 15216 | 2.30 |
| 5 | 30552 | 68437 | 2.24 | 34853 | 85655 | 2.46 |
| 6 | 75795 | 97092 | 1.28 | 92640 | 135941 | 1.47 |
| 7 | 15615 | 19551 | 1.25 | 20582 | 34354 | 1.67 |
| 8 | 6095 | 6549 | 1.07 | 9995 | 16059 | 1.61 |
| 9 | 40724 | 34916 | 0.86 | 55187 | 59119 | 1.07 |
| 11 | 2985 | 4767 | 1.60 | 4043 | 7410 | 1.83 |
| Total | 252443 | 440020 | 1.74 | 311014 | 622539 | 2.00 |

\* 1 - Mathematics and computer science, 2 - Physics, 3 - Chemistry, 4 - Earth sciences, 5 - Biology, 6 - Medicine, 7 - Agricultural and veterinary sciences, 8 - Civil engineering, 9 - Industrial and information engineering, 11 – Psychology

## 5.2 The statistical model

The simple increase in the number of self-citations in the years following the introduction of the ASN is not sufficient evidence that there has been a change in the self-citation behavior by academics. Testing the research hypotheses requires identification and control for other variables expected to predict the self-citation rates, as evidenced in the review of the literature, including:

1) *The number of publications the researcher publishes in a given year.* With the same propensity to self-cite, the number of self-citations per year will be greater with researcher's increasing production of publications in that year.
2) *The total number of publications authored by the researcher.* The expected relationship between number of self-citations in a given year and the size of publication portfolio up to that year is positive but non-linear, with decreasing increments.
3) *The field of research.* The intensity of citation varies across fields, therefore the analyses must be conducted at field level, classifying the professors in their CSs or SDSs.
4) *The cognitive proximity of publications in the publication portfolio.* The greater is the cognitive proximity (i.e. similarity of knowledge bases), the greater the likelihood that previous publications would have intellectual or cognitive influence on future ones, thereby legitimizing self-citations. Thus, all others equal, the more a scientist specializes (the less they diversify) the higher the expected number of self-citations, and vice versa.
5) *The citation impact of the self-cited publications.* If the works are highly cited by the scientific community, their authors can be equally expected to cite them.

Among the personal and social factors that could affect the benefits from increasing the self-citation rate, we should consider:

6) *Academic rank.* The intensity of citation varies across academic ranks. We expect the benefits from inflating self-citations to be less for full professors than for lower ranks.

7) *Gender*. It has been shown that, on average, males have a self-citation rate higher than females, although in Italy, we would expect greater incentive for women to increase their self-citations, since they have lower research performance.
8) *Territory*. The research performance of Italian professors varies by macro area, and from this we would expect varying incentive to inflate self-citations.
9) *The number of co-authors of the citing publications*. We could expect co-authors to exercise a kind of social control as a deterrent to opportunistic behavior, therefore the higher the number of co-authors, the smaller the opportunity for the researcher inappropriately self-cite.
10) *The number of international co-authors of the citing publications*. Social proximity could favor collusive behavior. The social distance at the international level is greater than at the domestic and intra-mural level, therefore offering less scope for opportunistic self-citing.
11) *The citation impact of the author's publication portfolio.* The less others mention the researcher's work, the greater would be the incentive for the researcher to inflate self-citations, to pass the ASN thresholds.

Variables 6-11 differ from 1-5 because they act on the benefits, risks and controls over opportunistic self-citations, and only indirectly on the number of self-citations. We select them on the assumption that there is a tendency towards opportunistic behavior by the scientific community, with the role in the statistical model of assessing any changes in such behavior among professors, controlling for potential benefits or risks.

We fit a Negative Binomial panel random effects model (Hilbe, 2011). Specifically, the response variable $y_{it}$ is the number of self-citations of professor $i$ in year $t$ (considering papers published in the latest 8 years),[14] assumed to be a negative binomial random variable with mean $\mu_{it}$ and over-dispersion parameter $\alpha$, so that the variance is $(1+\alpha\mu_{it})\mu_{it}$. The negative binomial is a more flexible model for count data than the Poisson, as it allows a variance larger than the mean. The exposure variable $n_{it}$ is taken to be the number of publications of professor $i$ in year $t$. Thus $\theta_{it} = \mu_{it}/n_{it}$ is the self-citation rate, for which we specify a log-linear regression model:

$$log\theta_{it} = \beta_1 x_{1it} + \beta_2 x_{2it} + \cdots + \beta_p x_{pit} + v_i \quad v_i \sim N(0,\sigma^2) \qquad [1]$$

The model has $p$ covariates, which may be time-varying, plus a random effect $v_i$ collecting time-constant unobserved factors of professor $i$. The regression coefficients are usually reported in exponentiated form, since $exp(\beta_k)$ is the rate ratio comparing two statistical units with a difference of 1 in the $k$-th covariate.

The model parameters are the $p$ regression coefficients plus the overdispersion-parameter $\alpha$ and the variance of the random effect $\sigma^2$. Estimation is performed with the "menbreg" command of Stata, which performs maximum likelihood with adaptive Gauss-Hermite quadrature for the random effect.

The target covariates, namely the covariates related to our research hypotheses, are:

---

[14] The bibliometric dataset extracted from the available I-NCR consists of the 2001-2017 scientific production of all Italian professors on staff in that period, thus for 2008 (lower end of the first period of analysis), the observation covers only the production over the previous eight years. This limit is not critical, since researchers tend to self-cite more recent works (Aksnes, 2003). We apply the same eight-year window to all years of both periods.

- *PostASN*: a time-varying dummy variable taking value 0 until 2012 and value 1 from 2013.
- *StatusASN*: a time-constant variable with four categories[15]
  - *Full professors*: do not participate in ASN, since they are already at top rank;
  - *Assistant and associate professors not accredited in ASN 2012*: professors who either did not participate or participated and failed, and would therefore be motivated to participate in future calls of ASN;
  - *Assistant professors accredited in ASN 2012*: having obtained accreditation for the role of associate professor, they would be motivated to participate in future calls of ASN, with the intention of obtaining accreditation for full professor;
  - *Associate professors accredited in ASN 2012*: having obtained accreditation for full professor, these have no need of further participation in ASN.

The categories of *StatusASN* have different incentives for self-citing. In the model, *PostASN* is interacted with *StatusASN* so that the change in self-citations after 2012 is estimated separately for each category of *StatusASN*.

The model includes the following control variables, all time-varying with the exception of gender:
- *Citable papers*: number of papers published in the last eight years.
- *Authors*: average number of authors of the papers published in current year.
- *International*: a dummy variable taking value 1 if there are international co-authors in the papers published in current year.
- *Diversification*: number of distinct WoS subject categories associated with the papers published in the last eight years.
- *UDA*: University disciplinary area of the professor in current year (10 dummy variables, with no reference category, since the intercept is omitted).
- *SDS*: Scientific disciplinary sector of the professor in current year (196 dummy variables, for each UDA the reference category is the first SDS).
- *Rank*: academic rank of the professor in current year: dummy variables for assistant, associate and full professor.
- *Macro-region*: geographical macro-region of the university of the professor in current year: dummy variables for the Italian North, Centre (reference category) and South macro-regions.
- *Female*: dummy variable taking value 1 for female professors.

The baseline self-citation rate is allowed to vary for each combination of rank and UDA, so there are 30 baseline rates.

As for other covariates evidenced in the review of the literature (specifically those indicated as 5 and 11 in the list above), we considered also using the average impact of self-cited papers in year *t*. However, this variable is determined simultaneously with the outcome of the model (the number of self-citations in year *t*), so its use as a covariate is questionable. Since the insertion of this variable in the final model does not lead to relevant changes in the other coefficients, we omit it.

---

[15] To build the variable *StatusASN*, we take the academic rank in 2012, just before ASN. If the professor has no record in 2012 due to zero self-citations, we take academic rank in the last observed year. Note that the dataset does not consider whether an assistant professor would be seeking accreditation for advancement to associate or full professor, however the latter case would be so rare that it can be ignored.

The exploratory analysis suggested the suitability of log-transformation of the quantitative variables (*Citable papers*, *Authors*, *Diversification*). For improvement in interpretability of the baseline rates, the log-transformed variables are then centered with respect to the sample means.

The calendar year is represented by an integer-valued variable taking values 0 (2008) to 9 (2017). This variable is inserted as a 4$^{th}$ order polynomial, for flexibility in modeling of the time pattern.

## 6. Results

First we present the descriptive statistics of the empirical analysis, then the inferential statistics, and finally show the degree of diffusion of the self-citing phenomenon.

### 6.1 Descriptive statistics

The descriptive statistics for the time-varying variables are reported in Table 4, and for the time-constant variables in Table 5.

*Table 4: Descriptive statistics for time-varying variables (120,615 professor-years)*

| Variable | Min | Max | Mean | Std Dev. |
|---|---|---|---|---|
| Citable papers | 1 | 614 | 31.009 | 34.433 |
| Authors | 1 | 1009 | 13.255 | 58.137 |
| Diversification | 1 | 70 | 11.557 | 6.771 |
| International | 0 | 1 | 0.560 | - |
| PostASN | 0 | 1 | 0.512 | - |
| Rank | | | | |
|    Assistant | 0 | 1 | 0.177 | - |
|    Associate | 0 | 1 | 0.413 | - |
|    Full | 0 | 1 | 0.410 | - |
| Macro-region | | | | |
|    North | 0 | 1 | 0.452 | - |
|    Centre | 0 | 1 | 0.279 | - |
|    South | 0 | 1 | 0.270 | - |

Table 4 shows that the continuous variables (*Citable papers*, *Authors* and *Diversification*) are right skewed. Indeed, they enter the model after a logarithmic transformation. For binary variables the mean is equivalent to the proportion, thus for *International* it turns out that in 56% of the observations there is at least one international co-author. Note that the mean of *PostASN* is 0.512, which is larger than 0.5 since there are slightly more observations after ASN. The values of *Rank* reveal that associate and full professors have nearly the same weight, whereas assistant professors are the smaller group. The variable *Macro-region* is defined on the basis of the affiliation in the year, for example 45.2% of the observations refer to professors affiliated to an institution in the North of Italy.

Table 5 shows that 28.7% of the professors are female and it displays the distribution of the key variable *StatusASN*, which combines the rank of the professor before ASN and the outcome of ASN.

*Table 5: Descriptive statistics for time-constant variables (15,037 professors)*

| Variable | Min | Max | Mean |
|---|---|---|---|
| Female | 0 | 1 | 0.287 |
| ASN status | | | |
|     Full professor | 0 | 1 | 0.377 |
|     Assistant or associate professor not accredited in ASN 2012 | 0 | 1 | 0.439 |
|     Assistant professor accredited in ASN 2012 | 0 | 1 | 0.062 |
|     Associate professor accredited in ASN 2012 | 0 | 1 | 0.122 |

## 6.2 Inferences from the statistical model

Table 6 reports the estimates of the parameters from the negative binomial panel random effects model [1] in terms of rate ratios, omitting those less relevant, i.e. the baseline rates (for each combination of academic rank and UDA), the coefficients of the 196 dummies for the SDS, and the 4 polynomial coefficients of the time pattern. The table also omits the over-dispersion parameter and the variance of the random effect, which we comment below.

The over-dispersion parameter α of the Negative Binomial is estimated to be 0.221 (s.e. 0.002), i.e. a substantial over-dispersion, meaning that the Poisson model would be inappropriate. The variance of the random effect $\sigma^2$, estimated at 0.353 (s.e. 0.006), due to the unobserved characteristics of the professors, is substantial, as is typical of panel data. In this case, a change of one standard deviation in the distribution of the random effect would entail a change in rate ratio of exp(√0.353)=1.812, i.e. considering two professors with the same covariates, the professor with random effect one standard deviation greater would have a self-citation rate 81.2% higher.

*Table 6: Estimates of the main parameters of the Negative Binomial panel random effect model for the analysis of the self-citation rate of Italian professors*

| Explanatory variable | Rate ratio | Std Err. | z | p-value | 95% Conf. Interval | |
|---|---|---|---|---|---|---|
| StatusASN | | | | | | |
|   Full professor (reference category) | 1.000 | - | - | - | - | - |
|   Assistant or associate professor not accredited in ASN 2012 | 0.884 | 0.019 | -5.8 | 0.000 | 0.848 | 0.921 |
|   Assistant professor accredited in ASN 2012 | 1.131 | 0.036 | 3.9 | 0.000 | 1.062 | 1.203 |
|   Associate professor accredited in ASN 2012 | 1.101 | 0.026 | 4.1 | 0.000 | 1.051 | 1.154 |
| PostASN | | | | | | |
|   Full professor | 1.099 | 0.014 | 7.4 | 0.000 | 1.072 | 1.127 |
|   Assistant or associate professor not accredited in ASN 2012 | 1.150 | 0.016 | 10.2 | 0.000 | 1.120 | 1.182 |
|   Assistant professor accredited in ASN 2012 | 1.042 | 0.020 | 2.2 | 0.029 | 1.004 | 1.082 |
|   Associate professor accredited in ASN 2012 | 1.033 | 0.015 | 2.2 | 0.030 | 1.003 | 1.064 |
| ln(Citable papers) | 2.304 | 0.019 | 100.2 | 0.000 | 2.267 | 2.342 |
| ln(Authors) | 0.928 | 0.005 | -13.3 | 0.000 | 0.918 | 0.939 |
| ln(Diversification) | 0.769 | 0.007 | -27.2 | 0.000 | 0.755 | 0.784 |
| International | 0.977 | 0.006 | -3.9 | 0.000 | 0.966 | 0.989 |
| North | 0.986 | 0.014 | -1.0 | 0.322 | 0.960 | 1.013 |
| Centre (reference category) | 1.000 | - | - | - | - | - |
| South | 1.022 | 0.016 | 1.4 | 0.168 | 0.991 | 1.053 |
| Female | 1.077 | 0.015 | 5.3 | 0.000 | 1.048 | 1.107 |

*Note: the standard error and the confidence interval refer to the rate ratio exp(β), while the z statistic and the corresponding p-value refer to the original parameter β.*

Table 6 shows the exponentiated coefficients exp(β), which are rate ratios. For example, the rate ratio 1.077 for *Female* means that considering two professors with identical values of covariates, except for gender, a female professor has a self-citation rate 7.7% higher than a male professor. Since the *z* statistic is equal to 5.3, this is statistically significant. Considering geographical area, the self-citation rate is slightly lower for professors affiliated with institutions in the North and higher for those in the South.

The control variables for self-citation behavior act in the expected directions: the self-citation rate increases with the number of citable papers and decreases with number of authors, degree of diversification, and presence of international authors. The rate ratios refer to an increase of 1 in the log scale, so it is difficult to appreciate the magnitude of the effects. To assist interpretation, we compute the rate ratios when the covariate increases from the first to the third quartile: the rate ratios are 2.617 for *Citable papers*, 0.944 for *Authors*, and 0.819 for *Diversification*. As expected, the number of citable papers is by far the most relevant control variable, although degree of diversification has remarkable effect.

The target variables, *StatusASN* (time-constant variable with four categories of status with respect to ASN) and *PostASN* (time-varying dummy variable for the years after ASN), are included with an interaction. The interpretation of the estimates is as follows:

- The coefficients of *StatusASN* concern self-citation rates of the four status categories prior to the first ASN, i.e. over the period 2008-2012. Compared to full professors (reference category), the assistant and associate professors who did not qualify in ASN had a rate ratio of 0.884, namely, adjusting for the control variables, their self-citation rate was 11.6% less; professors who did qualify in ASN had a higher rate ratio, specifically 1.131 for assistant professors and 1.101 for associate professors.
- The coefficients of *PostASN* measure the change after ASN for each category of *StatusASN*, i.e. comparing self-citation rates of 2013-2017 to those of 2008-2012. All categories show an increase, more so the assistant and associate professors not accredited in the 2012 ASN (1.150), followed by the full professors (1.099), the assistant professors who gained accreditation in 2012 (1.042), and the accredited associate professors (1.033).

Therefore, controlling for the available covariates, all professors increased their self-citation rates after the experience of the 2012 ASN. The magnitude of increase across categories of *StatusASN* seems to be related to each category's incentives for self-citing.

For further insight into the change after ASN, we fitted two alternative versions of the model. In the first version, *PostASN* is not interacted with *StatusASN*, obtaining a single coefficient measuring the overall change: the rate ratio results as 1.095 (s.e. 0.013), i.e., controlling for the available covariates, the overall self-citation rate increased 9.5% after 2012. In the second version, *PostASN* is further interacted with academic rank and UDA, obtaining the set of specific coefficients shown in Figure 1. Remarkably, despite some heterogeneity, the increase in the self-citation rate took place in every scientific area and for every rank.

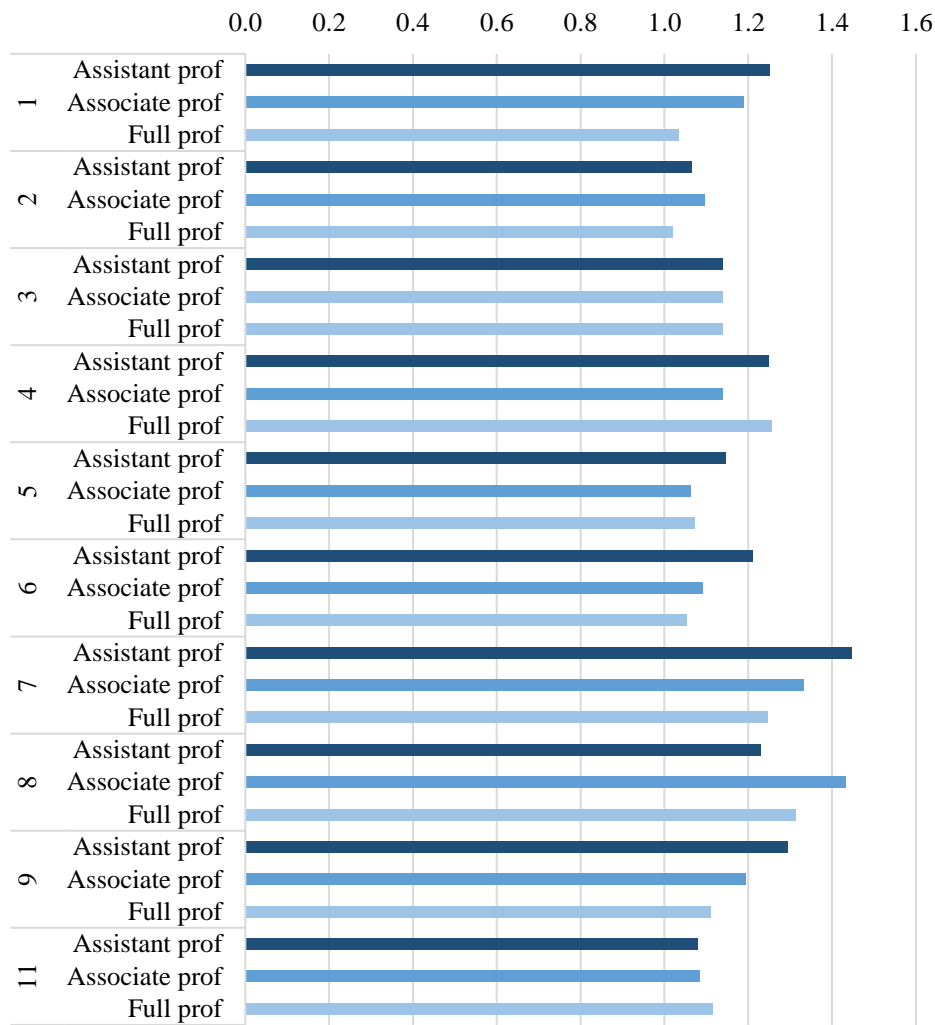

*Figure 1: Estimated self-citation rate by academic rank and UDA for full professors and for assistant and associate professors not accredited in ASN (all other covariates set at zero).*

A final consideration is that the model results rely on the dummy variable *PostASN*, which imposes a jump in the time pattern in 2013, immediately following the ASN. For an empirical check on whether this truly occurred, we explored the time pattern without imposing any structure, by fitting a model where the pattern is accounted by a set of year-specific dummy variables. Indeed, the estimates from that model show a notable jump in 2013. This is exemplified by Figure 2, which depicts the time pattern of the estimated self-citation rate of full professors in SDS FIS/01 (Experimental physics).[16]

---

[16] The choice of an SDS is required to compute the estimated rates, but note that the model assumes the same pattern across SDSs.

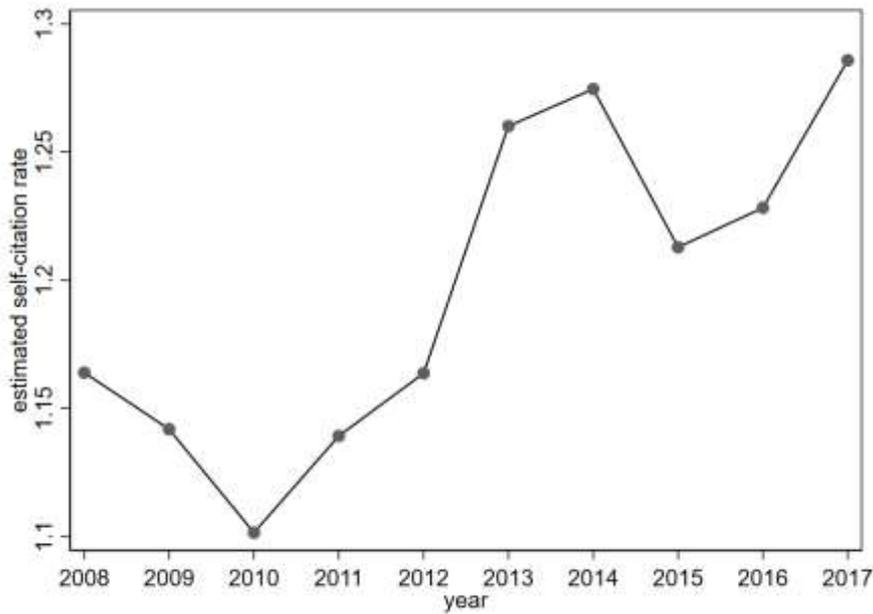

*Figure 2: Estimated self-citation rate by year (model with year-specific dummies), for full professors in SDS FIS/01 (Experimental physics) with all covariates set at zero*

**6.3 Diffusion of change in self-citation behavior**

The model-based analysis allowed us to establish that the increase in self-citations concerned all types of professors and all disciplines. However, the model parameters reflect an average behavior, and it would be interesting to evaluate the degree of heterogeneity in the behavior of the individual professors, and understand the effects of anomalous cases. To this end, we perform a descriptive analysis of the individual change, i.e. the difference between the mean self-citation rate before and after ASN. In this way, without adjusting for the control variables, we directly measure the change in the self-citation rate. This before-after difference cannot be evaluated for 778 of the 15037 professors, who have no self-citations before ASN. For the remaining 14259 professors, the mean difference is 0.251 with standard deviation 1.111. For the same professors, the mean self-citation rate was 1.201 before ASN, thus the average increase is 20.9%. This is remarkably larger than the model estimate as it is not adjusted for the covariates. The boxplots in Figure 3 show the distribution of the difference for the categories of *StatusASN*.

The distributions are similar: right skewed with several outliers, both positive and negative. Interestingly, positive outliers are only somewhat more than negative ones. If we exclude the extreme point at -11.302, the minimum is -6.000, at a distance of 6.251 from the mean 0.251. This leads us to question how many points would be beyond such a distance above the mean, namely how many points place beyond 0.251+6.251= 6.502. The answer is 22, i.e. only 22 professors out of 14259 (0.15%) show a high anomalous increase in their self-citation rate.

Of 14259 professors, a full 41.1% did not increase self-citation rates, and among the 58.9% who did there was much variability. The analysis of the outliers reveals only a few cases of exaggerated increase. From this, in general, there was no systematic increase in self-citation.

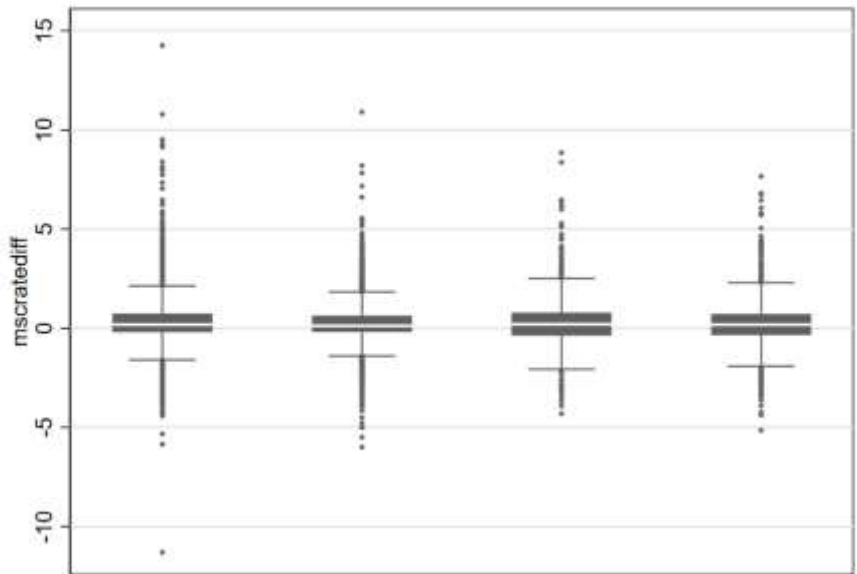

*Figure 3: Boxplots of the difference in the mean self-citation rate (2013-2017 vs 2008-2012). Left to right: full professors; assistant and associate professors not accredited in ASN; assistant professors accredited in ASN; associate professors accredited in ASN.*

## 7. Conclusions

Incentive systems aimed at pursuing continuous improvement are not free from side effects, including the alteration of behavior for purposes obtaining benefits. The phenomenon of self-citation gaming associated with evaluative scientometrics is a case in point. The analysis of the phenomenon is complex, since self-citation rates depend on a number of variables, which need to be controlled for when investigating the effects of incentive schemes.

We have shown that self-citation rates, i.e. the average number of self-cited works in the reference list of publications authored in a period of time, increases with the number of citable papers, while it decreases, in order of importance, with the number of authors, the cognitive distance of the citable publications (extent of diversification), and the presence of international authors.

Univariate analyses of the phenomenon, as often proposed in literature, are therefore purely descriptive, and insufficient to explain a cause-effect relationship between incentive schemes and possible opportunistic behavior. It would also be impossible to provide decision-makers with a deterministic indicator of self-citation rates and reference thresholds, for identification and potential sanction of dishonest researchers. Still, from a policy and strategic management perspective, it would be useful to understand if and to what extent the apparent increase in research performance, obtained through incentive schemes, is real or distorted by self-citation gaming. In the case that illegitimate behavior occurs, it would also be useful to understand whether gaming is heterogeneous among disciplines, academic ranks, genders or territories, for consideration of rescaling the performances accordingly, and for modulation of corrective actions.

For the purposes of ASN or similar incentive schemes, it is equally important to understand whether self-citation gaming is concentrated or widespread. Concentrated behavior would be more serious, risking that some would benefit to the detriment of others, while diffuse behavior would simply increase the values of the bibliometric thresholds in general, with no real benefit to anyone.

The results of this work show that Italian professors have responded to the ASN incentive with an average increase in self-citation rate of 9.5%. In practice, this means that after introduction of the ASN, all other covariates equal, an academic who had once self-cited an average of ten works in each new publication would now self-cite about eleven. We have observed that the increment is common to all disciplines and academic ranks, however it is sensitive to the relative incentive: highest for assistant and associate professors not accredited by the first ASN (+15%), followed by full professors (+9.9%), accredited assistant professors (+4.2%), and accredited associate professors (+3.3%). From this, it seems that following successful accreditation, the professors tend to relax their self-citation behavior.

Interestingly, the increase in self-citation following introduction of the ASN has not been widespread. Over 41% of professors have not increased their self-citation rates, which, for these, could raise questions of unfairness in the award of accreditation or financial considerations subject to citation-based evaluations. A future study could concern the performance analysis of this category, compared to those who have increased their self-citation rate.

As compared to the only multivariate study on the topic (Seeber, Cattaneo, Meoli, & Malighetti, 2019), our field of observation is much larger, in terms of number of observations (15,037 professors vs 866), SDSs (199 SDSs vs 4), and time window following the ASN (5 years vs 2). We specify an inferential model similar to Seeber's et al. 'Interaction' model, with the addition of a further status category, namely accredited professors in 2012. The main control variables are the same, while few others differ, which should not substantially change the findings, which are aligned.

The available data do not allow us to detect to what extent the observed increases are due to pure gaming by the researcher, i.e. illegitimate self-citing, or to greater care in legitimately crediting "own works". This could be the subject of specific case studies.

Certainly, not all the increment can be a priori condemned as gaming. Moreover, in spite of the apparently strong incentive for "gift authorship" practices, embedded in the VQR 2004-2010 national research assessment, our previous empirical analysis again found no evidence of opportunistic behavior by Italian universities (Abramo, D'Angelo, and Di Costa, 2019). We must also recall that before being published, all the works undergo the scrutiny of editors and reviewers, who should notice any excess and be able to discriminate legitimate self-citations from illegitimate ones.

We are all aware that incentive schemes are likely to induce a certain level of opportunistic behavior, and when critical evidence has appeared this has sometimes led to advocacy or suggestions of abandoning evaluation metrics and the relative incentive schemes (Baccini, De Nicolao, & Petrovich, 2019). In the specific case, before abandoning such systems, it would at least be necessary to assess whether, net of game playing, the schemes have indeed forwarded the aims of better selection of research personnel and increased productivity. Many studies are limited to examining only one side of the coin, i.e. opportunistic behavior. On the other side, in a previous work on the effectiveness of recruitment in Italian universities following introduction of the ASN, Abramo and D'Angelo (2020) show a sharp decline in the advance of unproductive scholars in the ranks of professors. To complete the picture, the authors intend to assess the impact of performance-based research funding on the research productivity of Italian professors.

Being suspicious of opportunistic behavior, another suggestion has been to exclude self-citations from evaluation metrics. However, this would penalize honest researchers, particularly those who publish a lot and operate in a highly specialized manner. From the current analysis, we can see that for the sake of excluding the one additional post-ASN self-citation, which might be illegitimate, we would then exclude a further ten, until now considered legitimate. Our personal position is that we should not attempt to correct opportunistic behavior through measures that would penalize those who never adopted it. Instead, it would make more sense for editors and reviewers to be observant of and inhibit illegitimate self-citations.

As concerns more general consideration of flaws or over-reliance on citation metrics for use in decision-making on recruitment, career advancement and research funding, our personal view is that, at least in Italy, the advantages of such metrics are under-exploited, and that where bibliometrics are used it is their design and application that should be corrected. Indeed, the evidence is that even the current citation-based metrics are superior to peer-review in predicting long-term impact of scientific publications (Abramo, D'Angelo, & Reale, 2019), and so the question becomes how to improve and wisely expand their use, rather than insist on returning to or retaining peer review of scientific production.

Concerning the improvement and application of the metrics, we note the comments of Heinze and Jappe (2020), comparing the use of bibliometric models in the Netherlands and Italian research systems: "The Netherlands follows a model of bibliometric professionalism, whereas Italy follows a centralized bureaucratic model that co-opts academic elites", and notwithstanding that "Italy has expert organizations with bibliometric competence, the design and implementation of the national evaluation exercise has remained outside their purview." For Italy, the challenge is one of drawing on the available knowledge to make the metrics as accurate as possible, then compiling them as carefully and systematically as possible, and using them in the best and fullest ways, without losing sight of their main limitations.

**Acknowledgements**
This research was funded by the Italian Ministry of Education, Universities and Research, BANDO PRIN 2017NKWYFC "The effects of evaluation on academic research: knowledge production and methodological issues", to Abramo, G. and D'Angelo, C.A.